# CONCEPT OF STAGED APPROACH FOR INTERNATIONAL FUSION MATERIALS IRRADIATION FACILITY

M. Sugimoto, M. Kinsho, H. Takeuchi, JAERI, Tokai, Ibaraki, Japan

*Abstract*

The intense neutron source for development of fusion materials planned by international collaboration makes a new step to clarify the technical issues for realizing the 40 MeV, 250 mA deuteron beam facility. The baseline concept employs two identical 125 mA linac modules whose beams are combined at the flowing lithium target. Recent work for reducing the cost loading concerns the staged deployment of the full irradiation capability in three steps. The Japanese activity about the design and development study about IFMIF accelerator in this year is presented and the schedule of next several years is overviewed.

## 1 INTRODUCTION

The International Fusion Materials Irradiation Facility (IFMIF) is an IEA collaboration to construct an intense neutron source for development of fusion materials [1]. The 250-mA, 40-MeV deuteron beam is required to satisfy the neutron flux level (wall load equivalent to 2 MW/m$^2$ ~ 9x10$^{13}$ neutrons/cm$^2$/s ~ 19 dpa/y for Fe) with enough irradiation volume (>500 cm$^3$). As the basic concept discussed during these five years of CDA (Conceptual Design Activity), a set of two identical 175 MHz, 125 mA linacs is employed to achieve the beam current requirement [2]. After the request from the Fusion Program Coordination Committee (FPCC) in January 1999, a plan with the reduction of the facility construction cost (estimated at 1996) and the project schedule with a staged approach to match to the fusion reactor development plan is proposed at 2000 FPCC meeting. It consists of three stages and each stage achieves 20%, 50% and 100% of the full irradiation capability shown above, respectively. The prospects for materials development are recognized though the series of research items: the selection of materials for ITER test blanket module as a near term milestone, the acquisition of engineering data for reactor prototype (like DEMO), and the evaluation of lifetime of candidate materials.

From the accelerator technology viewpoints, some essential key issues need to be solved before starting the construction, i.e. extremely stable 155 mA deuteron injector, 175 MHz coupled cavity cw-RFQ, precise beam dynamics simulation to realize the beam loss control, etc. The most problem should be addressed by prototyping, however, some prior verification about the component technology is necessary to initiate it. In the next several years, we concentrate on the restricted area of key component technologies to proceed to the next coming Engineering Validation Phase (EVP) as a preparation of construction phase.

## 2 STAGING CONCEPT

### 2.1 Overview

In the staged facility design, the layout of two 40 MeV deuteron linac modules becomes simple coplanar form to be upgraded easily. The major parameters of linac module are summarized in Table 1 and the layout of one accelerator module is shown in Fig. 1.

Table 1: Principal Parameters of Accelerator System

| Item | Specification | Description |
|---|---|---|
| Particle | D+ | H2+ for tests |
| No. of Modules | 1 or 2 | 1@ 1st/2nd stage |
| Beam Current | 50/125/250mA | 1st/2nd/3rd stage |
| Beam Energy | 32 and 40MeV | Selectable |
| Duty | 100% CW | Pulse for tests |
| Beam Size | 20cmWx5cmH | Uniform [1] |
| Energy Spread | )0.2MeV | Natural spread |
| RF Frequency | 175MHz | RFQ & DTL |
| RF Power | 9MW | 1MW unit x11 |
| Availability | > 88% | Scheduled op. |
| Maintainability | Hands-on | HEBT ends at target I/F valve |

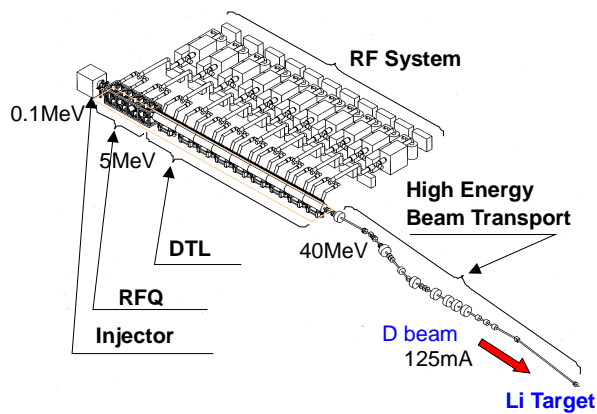

Figure 1: Layout of IFMIF accelerator module.

---

[1] Narrower width may be requested at 1st and 2nd stages to keep charge density of full current beam.

The number of irradiation test cells is reduced from two to one that would be possible to rearrange the schedule of irradiation tests, so that the High Energy Beam Transport has only one beam line. The electromagnetic pump is redesigned to minimize the volume of the loop for high-speed lithium flow used as the neutron-generating target. The resulting lithium inventory becomes 9 m$^3$ from prior value 21 m$^3$. The newly estimated cost indicates that the 1$^{st}$ phase of 50mA operation can be started by 38% of the total cost (~$800M) formerly obtained at CDA phase and the integral cost of all stages can be compressed to 60% of the former cost [3].

The construction/operation is divided into three stages: (1) 50 mA operation of a full performance linac for ~5 years, (2) full power 125 mA operation of the first linac for ~5 years, and (3) 250-mA operation with an addition of the second linac for more than 20 years. Other major parameters are not greatly changed from the CDA design but the cut of redundancy of the reduced cost design might influence to the overall availability, especially at the initial stage operation.

## 2.2 Injector

The ion source for 155mA deuteron beam with required quality is almost available at the present technology. Only the verification of long-term stability and long lifetime should be addressed, and these tasks will be performed in a couple of year. As the actual operation starts from 50mA in the staged approach, the lifetime issue is also relaxed.

On the other hand, LEBT is still problematic because of the less controllability of the space charge neutralization. The pulsing method to apply at the start up procedure is another unresolved issue. The use of H2+ beam at the prototype or commissioning phase brings the extra task to calibrate and correlate the measurements with D beam case.

## 2.3 RFQ

As shown in Fig. 1, the output energy of IFMIF RFQ is 5MeV (CDA design employed 8MeV output) and the final decision of the transition energy is made just before the construction probably. In any case, the length of RFQ exceeds 8 m and the coupled cavity technique developed by LANL [4] is needed to maintain the field uniformity along the structure. The beam loss in RFQ usually occurs at initial bunching section mainly and along the acceleration section in a small part, as shown in Fig. 2. The loss at low energy part will generate neutron due to D(d,n) reaction for the self-impinged deuteron at the vane surface. It may helpful to be coated by high-Z material at vane tip and to use a method of surface cleaning to remove deuteron gas periodically. For the loss at higher energy part high-Z material coating may also useful but the better solution is stop of RFQ with a small aperture size. This might push the lowering of transition energy because RFQ with large bore is inefficient accelerating structure. Again the final decision requires the acquisition of many precise calculations and accurate measurements.

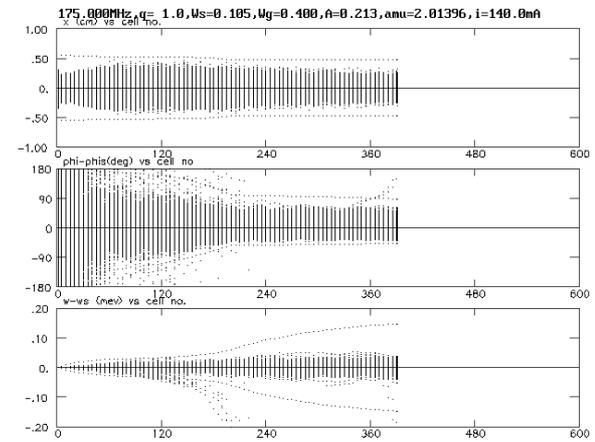

Figure 2: Typical result of particle tracking in IFMIF RFQ using PARMTEQ (top:x, middle:$\phi$-$\phi$s, bottom:W-Ws).

## 2.4 DTL

In the baseline design, Alvarez DTL is employed as the main accelerator, with single stem and post coupler. CDA design uses 3cm bore size for all drift tubes so that minimum incident energy is around 8 MeV if the conventional electromagnetic quadrupole using FoDo structure. The reduced cost design prefers the lower transition energy so that either the focusing scheme change like FoFoDoDo or bore radius change is necessary. Figure 3 indicates the PARMILA run of the former case. The resultant emittance growth is larger than that for FoDo case and we need to seek the best compromise on bore size. The gradient ramping at the beginning of DTL is another issue to be addressed at prototyping and extensive electromagnetic calculation is scheduled.

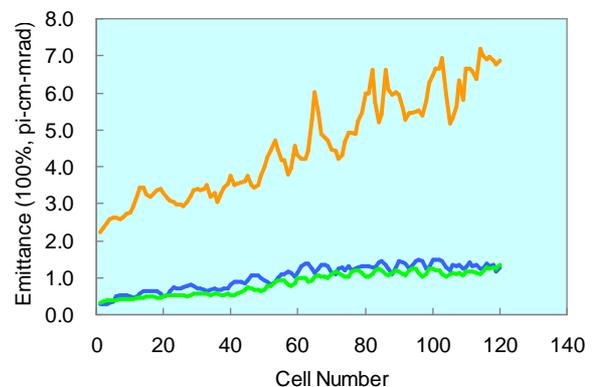

Figure 3: Emittance profile of IFMIF DTL by PARMILA.

## 2.5 RF System

The most of accelerator tanks is configured as multi-drive form using 2 independent 1MW RF amplifier units shown in Fig. 4. At the first stage 50 mA operation is achieved by removing one of two units and it is installed at the later stage. The circulator at the final output is not used in the current design because it might be error-prone component from the experience at ICRF heating. The serious analysis of such RF system control and response is necessary.

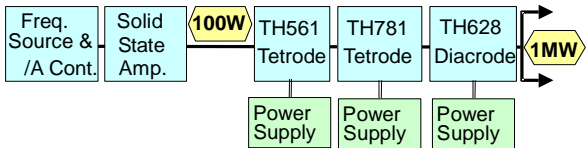

Figure 4: Layout of IFMIF RF power source unit.

## 2.6 HEBT

The design of new HEBT line is relaxed due to its simplicity of the beam transport line, however, it still requires the beam redistribution at target (20cm width and 5cm height with uniform distribution except ramping at both vertical ends). The resulting line consists of an achromatic parallel translation with two dipoles and static multipole magnet and imager qudrupoles as redistribution system and the last dipole bend, after that there is a 14m long drift space only till Li target.

The beam calibration dump placed at the middle of two Li target stations in CDA design is disappeared now, and alternative beam stop is desired for start up tuning purpose, which accepts several 100 kW power. The best place is straight end of the last dipole and it should be checked against the neutron back streaming from beam dump.

## 2.7 Superconducting Linac

From the beginning of the IFMIF design study, the superconducting linac (SCL) was considered as the promising alternative to DTL and the progress of general technology has been tracked. For the possible use in the future upgrade, the compatibility with DTL and SCL are always concerned. Fig. 5 shows the one of the low $\beta$ structure for IFMIF purpose.

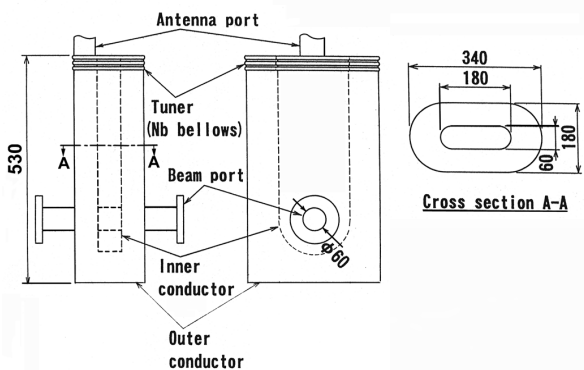

Figure 5: IFMIF SCAL quarter-wave structure [5].

## 3 DEVELOPMENT

The items covered in KEP include the long lifetime injector of accelerator system, the lithium flow stability test of target system, the temperature control of specimens of test cell facility, etc. The results of these tests contribute to realize the detailed design of the equipment for the next coming EVP to achieve the stable system operation. The items, injector test, RFQ cold model, DT packaging test are proposed as KEP tasks to be carried out in Japan with the possible international collaboration and the cooperative sharing between JAERI and the Japanese universities groups [6].

## 4 ADVANCED CONCEPT

The new scheme to realize the intense neutron source is a continuing task and a variation using Li flow without backwall us given in Fig.6, which is mixed with a partial energy recovery of deuteron beam to save electrical power.

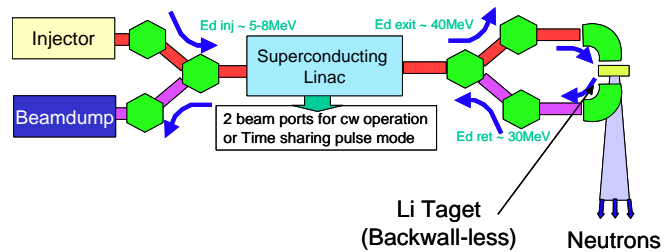

Figure 6:D-Li neutron source without backwall with partial deuteron beam energy recovery

## 5 SUMMARY

The materials development is one of the most important issues related to fusion programs, and it results in a new step to verify the key element technology, which needs to be carried out by using all possible international and domestic resources.